\documentclass[]{elsarticle}

\usepackage{lineno,hyperref}
\modulolinenumbers[5]

\usepackage[cmex10]{amsmath}
\usepackage{amsfonts}
\usepackage{nicefrac}
\usepackage{amssymb}
\usepackage{graphicx}
\usepackage{url}
\usepackage{epstopdf}
\usepackage{float}
\usepackage[T1]{fontenc}
\usepackage{ae,aecompl}
\usepackage{type1cm}
\usepackage{bbm}

\newtheorem{definition}{Definition}

\newtheorem{proposition}{Proposition}

\floatstyle{ruled}
\newfloat{algorithm}{tbp}{loa}
\providecommand{\algorithmname}{Algorithm}
\floatname{algorithm}{\protect\algorithmname}

\journal{Knowledge-Based Systems}
\begin{document}

\begin{frontmatter}

\title{From $t$-Closeness to Differential Privacy and Vice Versa in Data Anonymization}

\author{Josep Domingo-Ferrer}

\author{Jordi Soria-Comas}

\address{
Universitat Rovira i Virgili,
Dept. of Computer Engineering and Maths, 
UNESCO Chair in Data Privacy,
Av. Pa\"{\i}sos Catalans 26 - E-43007 Tarragona, Catalonia\\
\{josep.domingo,jordi.soria\}@urv.cat}

\begin{abstract}
$k$-Anonymity and $\varepsilon$-differential privacy are two mainstream
privacy models, the former introduced to anonymize
data sets and the latter to limit the knowledge
gain that results from including one individual in the data set.
Whereas basic $k$-anonymity
only protects against identity disclosure,
$t$-closeness was presented as an extension 
of $k$-anonymity that also protects against attribute disclosure.
We show here that, if not quite equivalent, $t$-closeness
and $\varepsilon$-differential privacy are strongly related to one another
when it comes to anonymizing data sets.
Specifically, $k$-anonymity for the quasi-identifiers combined
with $\varepsilon$-differential privacy for the confidential attributes 
yields stochastic $t$-closeness (an extension 
of $t$-closeness), with $t$ a function of $k$
and $\varepsilon$. Conversely,
$t$-closeness can yield $\varepsilon$-differential privacy 
when $t = \exp(\varepsilon/2)$ 
and the assumptions made by $t$-closeness about the prior and posterior
views of the data hold.

\end{abstract}


\end{frontmatter}


\section{Introduction}

$k$-Anonymity and $\varepsilon$-differential privacy are two mainstream
privacy models originated within the computer science community.
Their approaches
towards disclosure limitation are quite different: $k$-anonymity
is a model for releases of microdata ({\em i.e.}
individual records) that seeks to prevent record
re-identification by hiding each original record
within a group of $k$ indistinguishable
anonymized records, while $\varepsilon$-differential privacy 
seeks to limit the knowledge gain between data sets that differ in one individual.
Both models are often presented as antagonistic:
$\varepsilon$-differential privacy supporters view
$k$-anonymity as an old-fashioned privacy notion that offers only
poor disclosure limitation guarantees
(they argue that generating a 
a "noisy table" that safely provides accurate answers
to arbitrary queries is not feasible~\cite{Dwork11}),
while $\varepsilon$-differential
privacy detractors criticize the limited utility of $\varepsilon$-differentially
private outputs~\cite{Muralidhar10,Bambauer14}. 


In this paper we show that $t$-closeness (one of the extensions
of $k$-anonymity) and 
$\varepsilon$-differential privacy turn out to be strongly
related  
when it comes to anonymizing data sets.
Specifically, $k$-anonymizing the quasi-identifiers 
of a data set and ensuring 
$\varepsilon$-differential privacy for the confidential attributes yields
stochastic $t$-closeness (which in turn is an extension of $t$-closeness), 
with $t$ a function of
$\varepsilon$ and of the size of the $k$-anonymity equivalence classes.
Attaining standard $t$-closeness from $\varepsilon$-differential privacy 
is unfeasible because, being $\varepsilon$-differential 
privacy a stochastic mechanism, it is not possible to meet the requirements that
$t$-closeness puts on the empirical distribution of the confidential attributes.
To circumvent this issue, we introduce 
stochastic $t$-closeness, which is defined on the theoretical distribution
of the confidential attribute that results from the masking procedure (rather
than on a specific realization of it).
Conversely, $t$-closeness can yield guarantees {\em similar}
to $\varepsilon$-differential privacy for the confidential attribute. Attaining 
{\em exact} $\varepsilon$-differential privacy
from $t$-closeness is unfeasible because, for instance, $t$-closeness fails to provide any
protection if we assume that all the records except one are known by the intruder. However,
if the assumptions made by $t$-closeness about the prior and posterior views of the
data by the intruder hold, then the protection we get from $t$-closeness is equivalent 
to that of $\varepsilon$-differential privacy.

Section~\ref{sec2} contains background on $k$-anonymity,
$t$-closeness and $\varepsilon$-differential privacy.
Section~\ref{stochastic} introduces the concept of stochastic $t$-closeness.
Section~\ref{sec3} shows that 
combining $k$-anonymity and differential
privacy as sketched above yields (stochastic) $t$-closeness.
Section~\ref{sec4} shows that, under the assumptions 
made by $t$-closeness about the prior and posterior views of the data,
$t$-closeness provides differential privacy-like privacy guarantees for the 
confidential attributes.
A construction to attain $t$-closeness that relies on bucketization
is presented in Section~\ref{sec6}.
Section~\ref{related} reviews related work.
Conclusions and future research issues are summarized in
Section~\ref{sec7}.

The material in Section~\ref{sec6} was presented in the conference
paper~\cite{pst2013}.
The rest of sections in this paper are new.

\section{Background}
\label{sec2}

\subsection{$k$-Anonymity}

Assume a data set $X$ from which direct identifiers have been suppressed,
but which contains so-called {\em quasi-identifier} attributes, that is,
attributes ({\em e.g.} age, gender, nationality, etc.)
which can be used by an intruder to link records in $X$
with records in some external database containing direct identifiers.
The intruder's goal is to determine the identity
of the individuals to whom the values of
confidential attributes ({\em e.g.} health condition, salary, etc.)
in records in $X$ correspond
({\em identity disclosure}). See~\cite{Hund12} for further
details on disclosure attacks.

A data set $X$ is said to satisfy 
$k$-anonymity~\cite{SamaratiSweeney98,Samarati01} 
if each combination
of values of the quasi-identifier attributes in it is shared
by at least $k$ records.
We use the term {\em equivalence class} to refer to a maximal set of records that are indistinguishable 
with respect to the quasi-identifiers.
$k$-Anonymity protects against identity disclosure:
given an anonymized record in $X$,    
an intruder cannot determine the identity of  
the individual to whom the record (and hence  the confidential attribute values in it) corresponds. The reason is that there are at least $k$ records in $X$
sharing any combination of quasi-identifier attribute values.

The most usual computational procedure to attain $k$-anonymity
is generalization of the quasi-identifier attributes~\cite{Samarati01},
but an alternative approach is based on microaggregation of the
quasi-identifier attributes~\cite{Domi05}.

\subsection{$t$-Closeness}
\label{backt}

While $k$-anonymity protects against identity disclosure, as
mentioned above, it does not protect in general against 
{\em attribute disclosure},
that is,
disclosure of the value of a confidential attribute corresponding
to an external identified individual.
Let us assume a target individual $T$ for
whom the intruder knows the identity and the values of the 
quasi-identifier attributes. Let $G_T$ be a group of at least $k$
anonymized records sharing a combination of quasi-identifier
attribute values that is the only one compatible with $T$'s quasi-identifier
attribute values. Then the intruder knows that the anonymized 
record corresponding to $T$ belongs to $G_T$. Now,
if the values for one (or several) confidential attribute(s) 
in all records of $G_T$ are the same, the intruder learns
the values of that (those) attribute(s) for the target individual $T$.

The property of $l$-diversity~\cite{ldiver} has been proposed as an
extension of $k$-anonymity which tries to address the attribute
disclosure problem. A data set is said to satisfy $l$-diversity
if, 
for each equivalence class,
there are at least $l$ ``well-represented'' values for each
confidential attribute. Achieving $l$-diversity in general
implies more distortion than just achieving $k$-anonymity.
Yet, $l$-diversity may fail to protect against attribute
disclosure if the $l$ values of a confidential attribute
are very similar or are strongly skewed.
$p$-Sensitive $k$-anonymity~\cite{truta} is a property
similar to $l$-diversity, which shares similar shortcomings.
See~\cite{Domipsai08} for a summary of criticisms to
$l$-diversity and $p$-sensitive $k$-anonymity.

$t$-Closeness~\cite{tclose} is another extension of $k$-anonymity
which also tries to solve the attribute disclosure problem.
\begin{definition}
An equivalence class is said to satisfy $t$-closeness if the distance between the distribution of a sensitive attribute in this class and the distribution of the attribute in the whole table is no more than a threshold $t$. A table is said to satisfy $t$-closeness if all its equivalence classes satisfy $t$-closeness.
\end{definition}

$t$-Closeness clearly solves the
attribute disclosure vulnerability, although the original 
paper~\cite{tclose} did not propose a computational
procedure to achieve this property and did not mention
the large utility loss that achieving it is likely to
inflict on the original data.

\subsection{$\varepsilon$-Differential privacy}

Differential privacy was proposed by~\cite{Dwork06} as a relative privacy model
that limits the knowledge gain between data sets that differ in one individual.

\begin{definition}
\label{def:diff_priv}
A randomized function $\kappa$
gives $\varepsilon$-differential
privacy if, for all data sets $D_{1}$, $D_{2}$ such that one can
be obtained from the other by modifying a single record,
and all $S\subset Range(\kappa)$
\begin{equation}
\label{eqprob}
\Pr(\kappa(D_{1})\in S)\le\exp(\varepsilon)\times \Pr(\kappa(D_{2})\in S)
\end{equation}
\end{definition}

Originally the focus was on the interactive setting; that is, to protect the outcomes
of queries to a database. The assumption is that an anonymization mechanism
sits between the user submitting queries and the database answering them. 
The computational mechanism to attain $\varepsilon$-differential privacy
is often called $\varepsilon$-differentially private {\em sanitizer}. A usual sanitization approach
is noise addition: given a query $f$, the real value $f(D)$ is
computed, and a random noise, say $Y(D)$, is added to mask $f(D)$,
that is, a randomized
response $\kappa(D)= f(D) + Y(D)$ is returned. To generate
$Y(D)$, a common choice is to use a Laplace
distribution with zero mean and
$\Delta(X)/\varepsilon$ scale parameter, where:
\begin{itemize}
\item $\varepsilon$ is the differential privacy parameter;
\item $\Delta(f)$ is the $L_1$-sensitivity of $f$, that is,
the maximum variation of the query function between neighbor
data sets, {\em i.e.}, sets differing in at most one record.
\end{itemize}

Differential privacy was also developed for the non-interactive setting
in~\cite{Blum,Dwor09,Hardt2010,Chen11}. Even though a non-interactive data 
release
can be used to answer an arbitrarily large number of queries, in all
these proposals this feature is obtained at the cost of preserving utility only
for restricted classes of queries (typically count queries).
This contrasts with the general-purpose utility-preserving
data release offered by the $k$-anonymity model.

\section{Stochastic $t$-closeness}
\label{stochastic}

The fact that differential privacy is stochastic, while $t$-closeness is
deterministic, makes it impossible to guarantee that a differentially private
data set will satisfy $t$-closeness. To bridge 
this gap, we introduce the concept of {\em stochastic $t$-closeness}.

Let $X$ be a data set with quasi-identifier attributes collectively denoted
by $QI$ and confidential attributes $C_1, C_2, \cdots,C_n$. Let $N$ be the number
of records of $X$.

To attain $t$-closeness we need to partition $X$ into equivalence classes
such that the (empirical) distribution
of the confidential attributes over the entire 
$X$ is close to the 
(empirical) distribution of the confidential attributes 
within each of the equivalence classes: the distance
between the former distribution
and each of the latter distributions must be less than a threshold value $t$. 
If $c_1, c_2, \cdots,c_N$ are the values of a
confidential attribute in $X$, and 
$c_1, c_2, \cdots,c_{|E|}$ are the values 
of that attribute within
equivalence class $E$, the respective empirical 
distributions $\Pr(\cdot)$ and $\Pr_E(\cdot)$ can be computed as
\[ {\Pr}(B)= \frac{1}{N}\sum_{i=1, 2, \cdots,N} {\Pr}_{\mathbbm{1}_{c_i}} (B) 
\] and
\[ {\Pr}_{E}(B)= \frac{1}{|E|}\sum_{i=1, 2, \cdots,|E|} {\Pr}_{\mathbbm{1}_{c_i}}(B)
\]
where $B$ is any subset of values of the attribute,  
and $P_{\mathbbm{1}_{c_i}}(B)=1$ if 
and only if $c_i \in B$. 

The standard practice to generate a $t$-close data set $X'$ out of $X$
favors preserving the values of the confidential attributes within each
of the
equivalence classes. Probably the reason has to do with $t$-closeness being an extension
of $k$-anonymity. $k$-Anonymity is not concerned with the values of the confidential
attributes and, thus, attaining $k$-anonymity does not require any operation on them.
$t$-Closeness was initially attained by incorporating its constraints into $k$-anonymous data set generation algorithms; in this way, the values of the confidential attributes within each equivalence class were kept unmodified. 
In~\cite{tclose} and~\cite{tclose2}, the 
Incognito~\cite{incognito} and the Mondrian~\cite{mondrian} algorithms 
that were designed for $k$-anonymity 
are adapted for $t$-closeness. Modification of the confidential values was 
considered in~\cite{Rebollo2010}. Let $f$ be a function that 
masks the confidential 
attributes to bring the empirical distribution of the {\em modified} equivalence 
classes closer to the empirical distribution of the {\em modified} whole data set. 
Using the previous notations, the modified empirical distributions are computed as
\[ {\Pr}(B)= \frac{1}{N}\sum_{i=1, 2,\cdots,N} {\Pr}_{\mathbbm{1}_{f(c_i)}} (B) 
\] and
\[ {\Pr}_E(B)=\frac{1}{|E|}\sum_{i=1, 2,\cdots,|E|} {\Pr}_{\mathbbm{1}_{f(c_i)}}(B).
\]

The previous masking function $f$ is deterministic, but dealing with stochastic
masking functions would be interesting. Let $Z$ be a stochastic function. Attaining
$t$-closeness is not possible for $Z$, as the actual values depend on each
specific realization of $Z$. However, there is a natural extension of the concept
of $t$-closeness that deals smoothly with stochastic functions. Basically, we have
to stop working with empirical distributions, and use the distributions determined
by $Z$ instead. In terms of the formulas, we need to change $\mathbbm{1}_{f(\cdot)}$ 
by $Z(\cdot)$.

\begin{definition}
Let $X'$ be a data set obtained from $X$ by applying a stochastic function $Z$ to its confidential
attributes. We say that $X'$ satisfies stochastic $t$-closeness if, for each equivalence class $E$, the distance
between 
\[ {\Pr}(B)= \frac{1}{N}\sum_{i=1,2, \cdots,N} {\Pr}_{Z(c_i)} (B) 
\] and
\[ {\Pr}_E(B)= \frac{1}{|E|}\sum_{i=1, 2, \cdots,|E|} {\Pr}_{Z(c_i)}(B)
\]
is less than a threshold $t$.
\end{definition} 

Although a deeper analysis of stochastic $t$-closeness might be interesting, our aim here is
to relate it to differential privacy and the above 
definition will suffice to this aim.

\section{From differential privacy to (stochastic) $t$-closeness}
\label{sec3}

This section develops one of the main results in this paper: the implication
between $\varepsilon$-differential privacy and stochastic $t$-closeness.
Before going into that result, we need to deal with an important aspect
of $t$-closeness: the distance between the probability distributions used.

Stochastic $t$-closeness, just as $t$-closeness, is about making 
the distance between the data set-level and the equivalence class-level
distribution of the confidential attribute less than a threshold value $t$
for any equivalence class. 
When $t$-closeness was introduced, the Earth Mover's distance (EMD)
was proposed~\cite{tclose}. 
The EMD measures the minimal amount of work required
to transform one distribution into another by moving probability mass
between each other. 

The specific distance used can actually 
be viewed as an additional parameter of $t$-closeness
and it  has a significant impact on the kind of privacy guarantees offered.
As we aim to show that differential privacy for the confidential attributes
leads to stochastic $t$-closeness, we take a distance function that 
is more suited to differential privacy.

\begin{definition}
\label{def:distance}Given two random distributions $\mathcal{D}_{1}$
and $\mathcal{D}_{2}$, we define the distance between $\mathcal{D}_{1}$
and $\mathcal{D}_{2}$ as:
\[
d(\mathcal{D}_{1},\mathcal{D}_{2})=
\max_{S}\left\{\frac{\Pr_{\mathcal{D}_{1}}(S)}{\Pr_{\mathcal{D}_{2}}(S)},
\frac{\Pr_{\mathcal{D}_{2}}(S)}{\Pr_{\mathcal{D}_{1}}(S)}\right\}
\]
where $S$ is an arbitrary (measurable) set, and we take the quotients
of probabilities to be zero, if both $\Pr_{\mathcal{D}_{1}}(S)$ and
$\Pr_{\mathcal{D}_{2}}(S)$ are zero, and to be infinity if only the
denominator is zero.
\end{definition}

We are ready to move to the main result of the section: the implication between differential
privacy and $t$-closeness.

\begin{proposition}
\label{prop1}
Let $X$ be an original data set and $X'$ be a corresponding anonymized data set such that 
the projection of $X$ on the confidential attributes 
is $\varepsilon$-differentially private. Then $X$ satisfies stochastic $t$-closeness with
\[t = \max_E  \frac{|E|}{N}\left(1+\frac{N-|E|-1}{|E|}\exp(\varepsilon)\right) \]
where $E$ is an equivalence class.
\end{proposition}

{\bf Proof}.
Let $E$ be an equivalence class of $X$ and let $c_1, c_2,\cdots,c_{|E|}$ be the values of 
the projection of $X$ on the confidential attributes.
For $t$-closeness to hold, it must be $\Pr_E(B) \le t \times \Pr(B)$ and $\Pr(B) \le t \times \Pr_E(B)$, for any $B$.

We start by deriving an inequality that will later be used. 
If $Z$ satisfies $\varepsilon$-differential privacy, we have $\exp(-\varepsilon)\times \Pr_{Z(c_j)}(B) \le \Pr_{Z(c_i)}(B)\le \exp(\varepsilon)\times \Pr_{Z(c_j)}(B)$, for all $i$ and $j$. In particular, we have 
\[\frac{\exp(-\varepsilon)}{|E|} \sum_{j=1, 2,\cdots,|E|} {\Pr}_{Z(c_j)}(B) 
   \le {\Pr}_{Z(c_i)}(B) \]
\[
   \le \frac{\exp(\varepsilon)}{|E|} \sum_{j=1, 2, \cdots,|E|} {\Pr}_{Z(c_j)}(B). \]
Let $d = \sum_{i=1, 2, \cdots,|E|} {\Pr}_{Z(c_i)}(B)$. The previous equation becomes $\frac{1}{|E|}\exp(-\varepsilon) d 
   \le {\Pr}_Z(c_i) 
   \le \frac{1}{|E|}\exp(\varepsilon) d$.
By taking the sum for $i$ in $|E|+1, |E|+2, \cdots,N$, we have
\begin{equation} \label{eq:ine1}
\frac{N-|E|}{|E|}\exp(-\varepsilon)d 
  \le \sum_{i=|E|+1, |E|+2, \cdots,N} {\Pr}_{Z(c_i)}(B); 
\end{equation}
\begin{equation}  \label{eq:ine2}
\sum_{i=|E|+1, |E|+2, \cdots,N} {\Pr}_{Z(c_i)}(B)
  \le \frac{N-|E|}{|E|} \exp(\varepsilon)d. 
\end{equation}

Now we turn to the $t$-closeness requirements. We start with ${\Pr}_E(B) \le t \times {\Pr}(B)$. 
This inequality can be rewritten as
\[ \frac{1}{|E|} \sum_{i=1, 2, \cdots,|E|} {\Pr}_{Z(c_i)}(B) 
   \le
   t \times \frac{1}{N} \sum_{i=1, 2, \cdots,N} {\Pr}_{Z(c_i)}(B).
\]
In terms of $d$ it becomes
\[ \frac{1}{|E|} \times d \le t \frac{1}{N} \left(d + \sum_{i=|E|+1,
|E|+2, \cdots,N} {\Pr}_{Z(c_i)}(B)\right).
\]
By Inequality (\ref{eq:ine1}) the following inequality implies the previous one
\[ \frac{1}{|E|} \times d \le t \frac{1}{N} \left(d + \frac{N-|E|}{|E|}\exp(-\varepsilon) d  \right)
\]

By operating on the previous inequality we get
\[t \ge \frac{N}{|E|}\left(1+\frac{N-|E|}{|E|}\exp(-\varepsilon)\right)^{-1}.
\]

By following a similar process with inequality $P(B) \le t \times P_E(B)$
we conclude
\[t \ge \frac{|E|}{N}\left(1+\frac{N-|E|}{|E|}\exp(\varepsilon)\right).
\]

The conclusion is that if $Z$ satisfies $\varepsilon$-differential privacy, then we get 
$t$-closeness on $X'$ for
\[t = \max_E\{\frac{N}{|E|}\left(1+\frac{N-|E|}{|E|}\exp(-\varepsilon)\right)^{-1}, \]
\[ \frac{|E|}{N}\left(1+\frac{N-|E|}{|E|}\exp(\varepsilon)\right) \}.
\] 

It can be seen that the second term is always greater than the first one, which leads to
\[t = \max_E  \frac{|E|}{N}\left(1+\frac{N-|E|-1}{|E|}\exp(\varepsilon)\right). \]
\hfill $\Box$

There are two approaches to enforce $t$-closeness for 
a data set that contains multiple 
confidential attributes: (i) take the confidential attributes together as a single confidential 
attribute and seek $t$-closeness for it using the joint distribution, and (ii) deal with each
confidential attribute separately, that is,
seek $t$-closeness independently for each confidential attribute. 
The first approach
is stronger in terms of privacy guarantees, and it is 
the one we have used in the
previous proposition. If instead of having $\varepsilon$-differential privacy for the projection
of the data set on the confidential attributes we had $\varepsilon$-differential privacy for each
individual attribute, then the second approach would be more suitable.

Proposition~\ref{prop1} shows that $\varepsilon$-differential privacy for the confidential attributes
implies (stochastic) $t$-closeness for a $t$ that depends on the number of records, the cardinality of
the equivalence classes and $\varepsilon$. The proposition shows that $\varepsilon$-differential
privacy is stronger than (stochastic) $t$-closeness as a privacy model. From a more practical point
of view, Proposition~\ref{prop1} can be seen as a possible approach to generate a $t$-close data set.
The confidential attribute of such a data set has not only the privacy guarantees provided by 
$t$-closeness but also the ones given by $\varepsilon$-differential privacy.

\section{$\varepsilon$-Differential privacy through $t$-closeness}
\label{sec4}

In this section we show that 
if the conditions under 
which $t$-closeness provides its privacy guarantee hold
for a data set,
then we have differential privacy on the projection
over the confidential attributes. 

The quasi-identifier attributes are excluded from our discussion. 
The reason is that $t$-closeness offers no
additional protection to the quasi-identifiers beyond what $k$-anonymity does. 
For example, we may learn that an individual is not in the data set if there 
is no equivalence class in the released $t$-close data whose quasi-identifier
values are compatible with the individual's.

The main requirement for the implication between 
$t$-closeness and differential privacy 
relates to the satisfaction of the $t$-closeness requirements 
about the prior and posterior
knowledge of an observer. $t$-Closeness assumes 
that the distribution of the confidential
data is public information (this is the prior view of observers about the confidential
data) and limits the knowledge gain between the 
prior and posterior view (the distribution of
the confidential data within the equivalence classes) 
by limiting the distance between 
both distributions.

Similarly to Section~\ref{sec3}, to make $t$-closeness 
closer in meaning to differential privacy, we use the distance proposed in 
Definition~\ref{def:distance}.
If the distributions $\mathcal{D}_{1}$ and $\mathcal{D}_{2}$ of
Definition~\ref{def:distance} are discrete (as is the case for 
the empirical distribution of a confidential
attribute in a microdata set), computing the distance between them
is simpler: taking the maximum over the possible individual values
suffices.

\begin{proposition}
\label{prop1}
If distributions $\mathcal{D}_{1}$ and $\mathcal{D}_{2}$ take values
in a discrete set $\{x_{1}, x_{2}$ $\cdots,$ $x_{N}\}$, then the distance $d(\mathcal{D}_{1},\mathcal{D}_{2})$
can be computed as 
\begin{equation}
d(\mathcal{D}_{1},\mathcal{D}_{2})=\max_{i=1, 2,\cdots,N}\left\{\frac{\Pr_{\mathcal{D}_{1}}(x_{i})}{\Pr_{\mathcal{D}_{2}}(x_{i})},\frac{\Pr_{\mathcal{D}_{2}}(x_{i})}{\Pr_{\mathcal{D}_{1}}(x_{i})}\right\}\label{eq:distance}.
\end{equation}
\end{proposition}

Suppose that $t$-closeness holds; that is, the data set $X'$ consists 
of several equivalence classes selected in such a way that the multiplicative 
distance proposed in Definition~\ref{def:distance} between the distribution of 
the confidential attribute over the whole data set and 
the distribution within each of 
the equivalence classes is less than $t$. We will show that, if the assumption
on the prior and posterior views of the data made by $t$-closeness holds, 
then $\exp(\varepsilon/2)$-closeness implies 
$\varepsilon$-differential privacy. 
A microdata release can
be viewed as the collected answers to a set of queries, where each query
requests the attribute values associated
to a different individual. As the queries relate to different individuals,
checking 
that differential privacy holds for each
individual query suffices, by parallel composition, to check 
that it holds for entire data set.
Let $I$ be a specific individual in the data set. 
For differential privacy to hold for the query associated to individual $I$, 
including $I$'s data in 
the data set vs not including them must modify the probability
of the output by a factor not greater than $\exp (\varepsilon)$, where $\varepsilon$
is differential privacy parameter.

\begin{proposition}
\label{prop3}
Let $k_I(D)$ be the function that returns the view on $I$'s confidential data given $D$.
If the assumptions required for $t$-closeness to provide a strong privacy guarantee hold, 
then $\exp(\varepsilon/2)$-closeness on $D$ implies $\varepsilon$-differential privacy of $k_I$.
In other words, if we restrict the domain of $k_I$ to $\exp(\varepsilon/2)$-close data sets, then  
we have $\varepsilon$-differential privacy for $k_I$.
\end{proposition}

{ \bf Proof}.
Let $D_1$ and $D_2$ be two data sets satisfying $\exp(\varepsilon/2)$-closeness for the combination
of all confidential attributes.
For $\varepsilon$-differential privacy to hold, we need
$\Pr(k_I(D_1)\in S)\le \exp(\varepsilon) \times \Pr(k_I(D_2)\in S)$.
We consider four different cases: (i) $I\notin D_1$ and $I\notin D_2$, (ii) $I\notin D_1$
and  $I\in D_2$, (iii) $I\in D_1$ and  $I\notin D_2$, and (iv) $I\in D_1$ and $I\in D_2$. 

In case (i), the posterior view does not provide information about $I$ beyond 
the one in the prior view: we have $k_I(D_1)=k_I(D_2)$. Hence, 
$\varepsilon$-differential privacy is satisfied.

Cases (ii) and (iii) are symmetric. We focus on case (ii). Given the
 assumptions, we
have that $k_I(D_1)$ has the distribution of the confidential attribute on the whole table,
while $k_I(D_2)$ has the distribution of the confidential attribute of the equivalence
class that contains $I$. Because of $\exp (\varepsilon/2)$-closeness, the distributions of
$k_I(D_1)$ and $k_I(D_2)$ differ by a factor not greater than $\exp (\varepsilon/2)$ and,
therefore, satisfy $\varepsilon$-differential privacy.

Case (iv) is a composition of the previous case for $k_I(D_1)$ and $k_I(D_2)$. Both 
$k_I(D_1)$ and $k_I(D_2)$ differ from the distribution of the confidential attribute on 
the whole table by a factor of $\exp (\varepsilon/2)$. Thus, they differ from each other
by a factor not greater than $\exp (\varepsilon)$, as we wanted.
\hfill $\Box$

Parallel to what we did in Section~\ref{sec3}, 
the previous proposition deals with all
confidential attributes at once. That is, we assume that $t$-closeness holds for the
combination of all the confidential attributes and we see that we get differential privacy
for the same combination of attributes.
 If $t$-closeness is satisfied only for a specific confidential attribute
(or subset of confidential attributes),
then we get differential privacy for that
attribute (or subset of confidential attributes).

Proposition~\ref{prop3} shows that, if the assumptions of $t$-closeness about the prior
and posterior views of the intruder are satisfied, then the level of disclosure risk 
limitation provided by $t$-closeness is as good as the one of $\varepsilon$-differential privacy.
Of course, differential privacy is independent of the prior knowledge, so Proposition~\ref{prop3}
does not apply in general. However, when it applies, it provides an effective way of
generating an $\varepsilon$-differentially private data set.

\section{A bucketization construction to attain $t$-closeness}
\label{sec6}

We want to reach $t$-closeness using the distance of Definition~\ref{def:distance}. 
A problem we face is that $t$-closeness is not attainable if there are confidential 
attribute values with multiplicity less than the number of equivalence classes. The 
reason is that, when an equivalence class lacks one of the values of the confidential 
attribute, the distance between the equivalence class-level distribution and the data
set-level distribution is infinity (according to Definition~\ref{def:distance}). To
circumvent this issue, instead of working with the empirical distribution of the 
confidential attribute, we work with a bucketized version of it, where points are
clustered into a set of buckets $B_1, B_2, \cdots,B_n$.

\begin{figure}
\begin{centering}
\includegraphics[width=8.5cm]{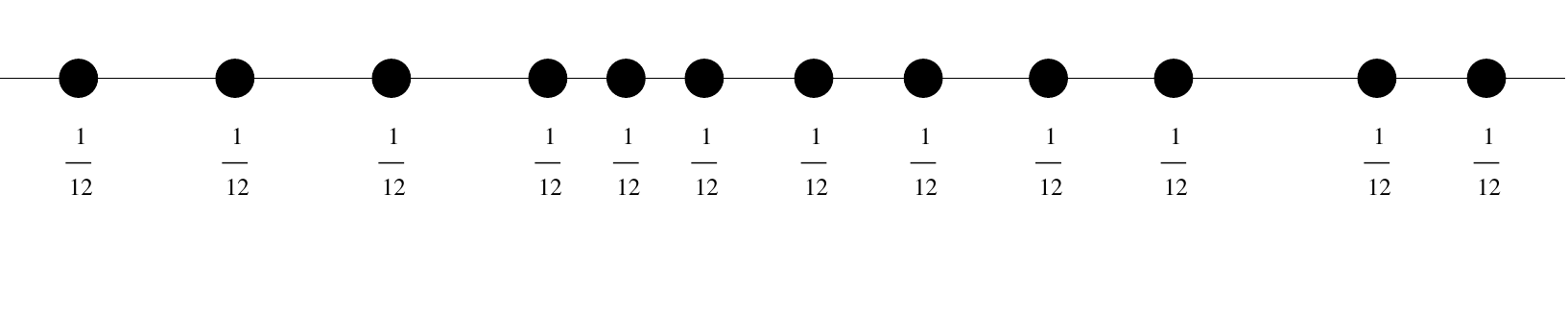}
\par\end{centering}

\begin{centering}
\includegraphics[width=8.5cm]{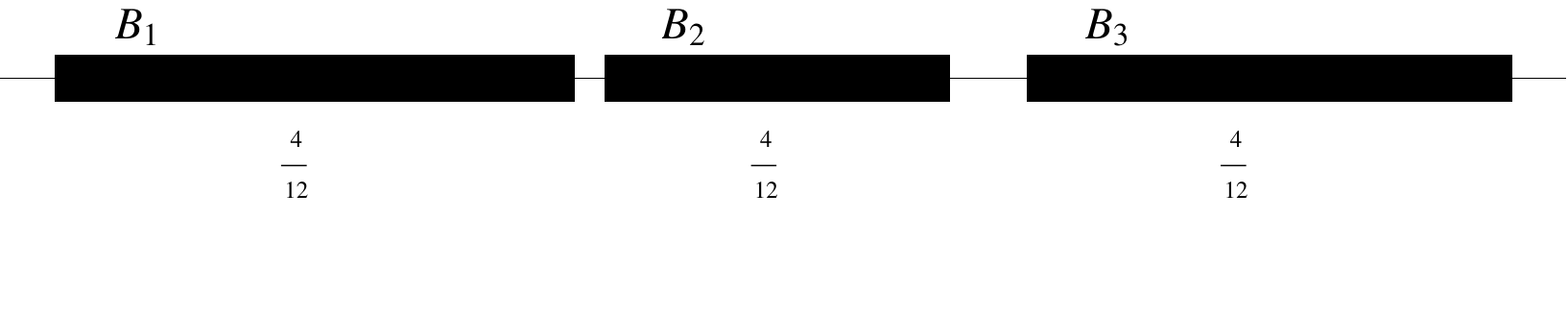}
\par\end{centering}

\caption{\label{fig:discretization}Top, original confidential attribute 
values. Bottom, bucketized confidential attribute values.}
\end{figure}

Before going into the details, we give an overview of the steps required to attain
$t$-closeness. Figure~\ref{fig:discretization} shows two sets of data. At the top 
there are the original values of the confidential attribute. According to the previous
discussion, $t$-closeness (for a finite $t$) is not feasible for it. At the bottom, 
there is a version of the confidential attribute where data have been clustered in 
buckets $B_1$, $B_2$ and $B_3$ that contain four points each. The 
granularity reduction 
of the data makes it feasible to attain $t$-closeness for a finite $t$.

\begin{figure}
\begin{centering}
\begin{tabular}{cc}
\includegraphics[width=6cm]{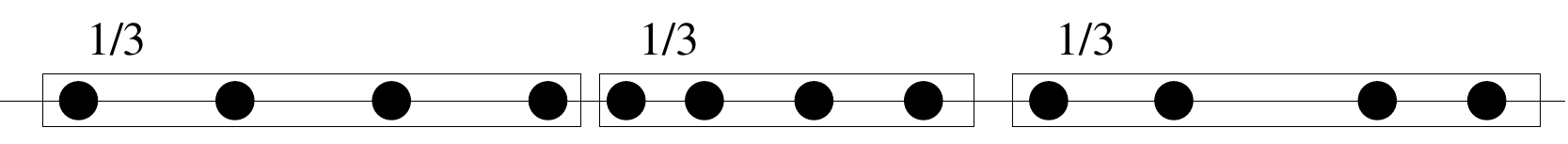} & original data\tabularnewline
\includegraphics[width=6cm]{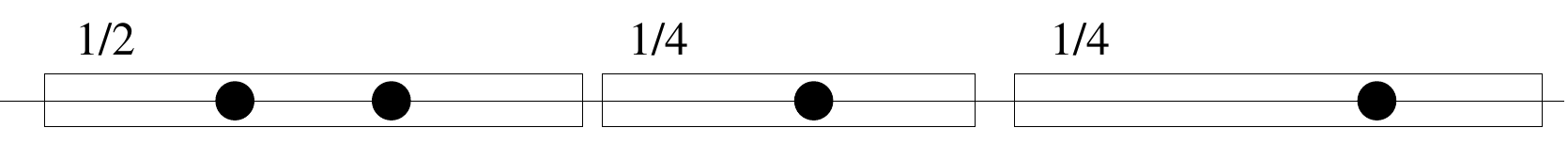} & equivalence class $E_{1}$\tabularnewline
\includegraphics[width=6cm]{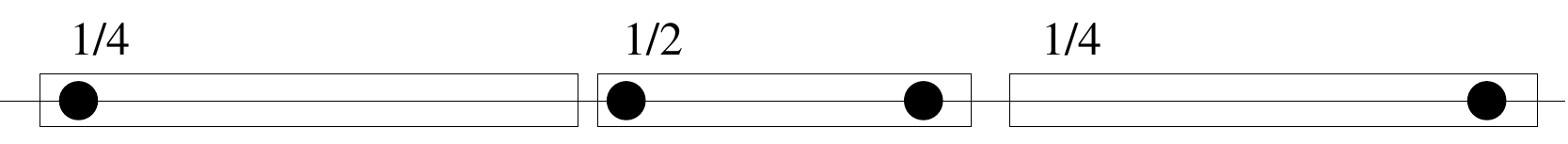} & equivalence class $E_{2}$\tabularnewline
\includegraphics[width=6cm]{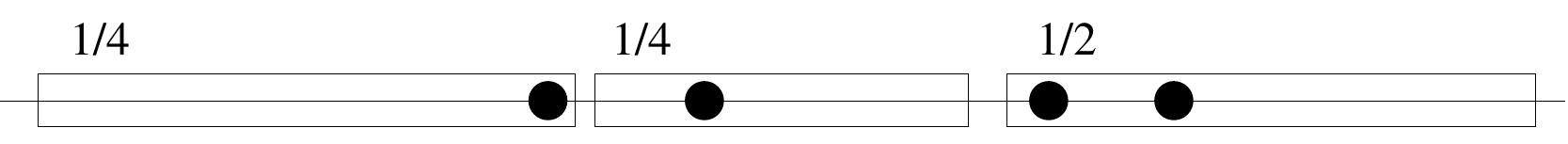} & equivalence class $E_{3}$\tabularnewline
\end{tabular}
\par\end{centering}
\caption{\label{fig:t_close_partition}Sample partition that satisfies $1.5$-closeness}
\end{figure}

For instance, Figure~\ref{fig:t_close_partition} shows a partition of the data set in 
equivalence classes that satisfies $1.5$-closeness, according to the previously defined 
distance. The empirical distribution of the original data assigns probability $1/3$ to each 
of the buckets $B_1$, $B_2$ and $B_3$; hence, the bucket-level distribution $\mathcal{D}$ 
of the confidential attribute in the original data set is $\Pr(B_1)=\Pr(B_2)=\Pr(B_3)=1/3$.
Each of the equivalence classes in the partition ($E_{1}$, $E_{2}$, $E_{3}$) takes 
either one or two points from each bucket. Thus, the bucket-level empirical distribution 
$\mathcal{D}(E_1)$ of the confidential attribute for equivalence class $E_1$ is $\Pr(B_1)=1/2$
and $\Pr(B_2)=\Pr(B_3)=1/4$; for equivalence class $E_2$, the distribution, denoted by 
$\mathcal{D}(E_2)$, is $\Pr(B_1)=\Pr(B_3)=1/4$ and $\Pr(B_2)=1/2$; for equivalence class 
$E_3$, the distribution, denoted by $\mathcal{D}(E_3)$, is $\Pr(B_1)=\Pr(B_2)=1/4$ and 
$\Pr(B_3)=1/2$. By using Equation~(\ref{eq:distance}) to measure the distance between 
$\mathcal{D}$ and $\mathcal{D}(E_i)$, for all $i$, we conclude that the generated partition 
satisfies $1.5$-closeness.
In Figure~\ref{fig:t_close_partition} we have depicted both the original set of values of 
the confidential attribute and the generated buckets.

The rest of the section focuses on determining the conditions that the bucketization and the
partition in equivalence classes must satisfy to maximize data utility. To fix notations,
we assume a data set $X$ of size $N$. The granularity of the confidential attribute is reduced
by considering
 $b$ buckets $B_1, B_2, \cdots,B_b$, each of them containing $b_i$ 
original values (that is, each of them having probability $b_i/N$). 
The partition in equivalence classes consists of $e$ equivalence classes 
$E_1, E_2, \cdots,E_e$, 
each of them containing $e_i$ (bucketized) records. Let $k$ be the minimum among the $e_i$'s.

The necessary condition, discussed at the beginning of the section, for $t$-closeness to be 
attainable can be reformulated according to the stated notation as $b_i \ge e$ for all $i$.

\subsection{Optimal bucketization}
\label{sub:Bucketizing}
The selected bucketization of the confidential attribute has a large impact on data utility.
To minimize the damage to data utility we should generate buckets that are as homogeneous 
as possible. In general, if a distance function is used to measure the similarity of the values 
of the confidential attribute, a clustering algorithm can be used to generate homogeneous buckets. 
The size of the buckets is a parameter that can have a large impact on bucket homogeneity: the
smaller the buckets, the more homogeneous they can be. However, taking
too small  
buckets may defeat the purpose of the bucketization,
that is, reducing the granularity of
the confidential attribute so that $t$-closeness is attainable. 
In this section we seek to 
determine the optimal bucket size.

\begin{figure}
\begin{centering}
\includegraphics[width=6cm]{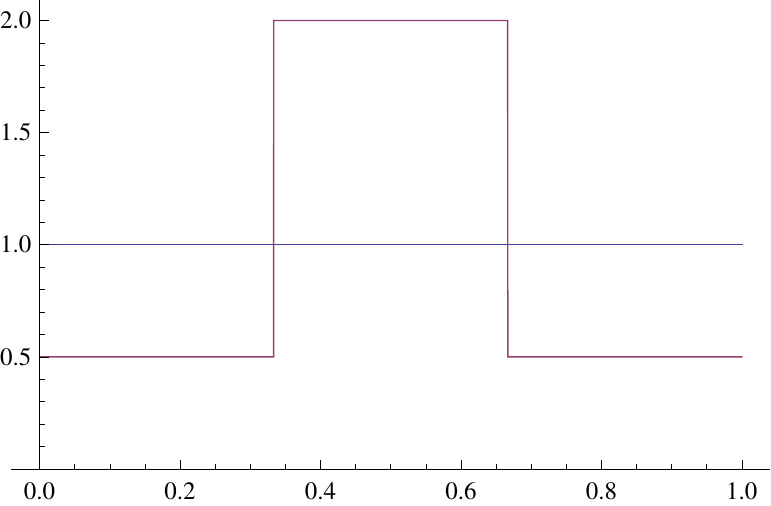}
\par\end{centering}

\caption{\label{fig:2_close_distribution}Probability distributions satisfying
$2$-closeness with the distance of Definition~\ref{def:distance}}
\end{figure}

Figure~\ref{fig:2_close_distribution} illustrates two probability distributions:
the uniform distribution represents the global distribution of the confidential attribute 
(over the whole data set), and the other distribution corresponds to the confidential attribute 
restricted to an equivalence class $E_i$. These two distributions satisfy $2$-closeness with the 
distance of Definition~\ref{def:distance}: the density of the restriction to $E_{i}$ is $1/2$ 
for the entire range of values of the confidential attribute, 
except for a subrange where it is 2. 

\begin{figure*}
\begin{center}
\begin{tabular}{ccc}
\includegraphics[width=5cm]{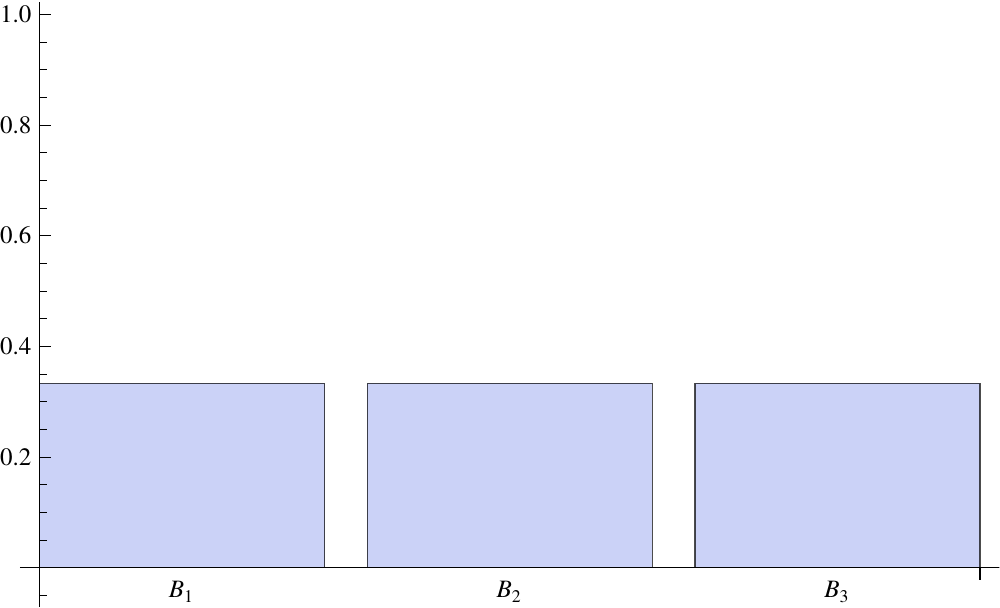}
& \rule{0.5cm}{0cm} &
\includegraphics[width=5cm]{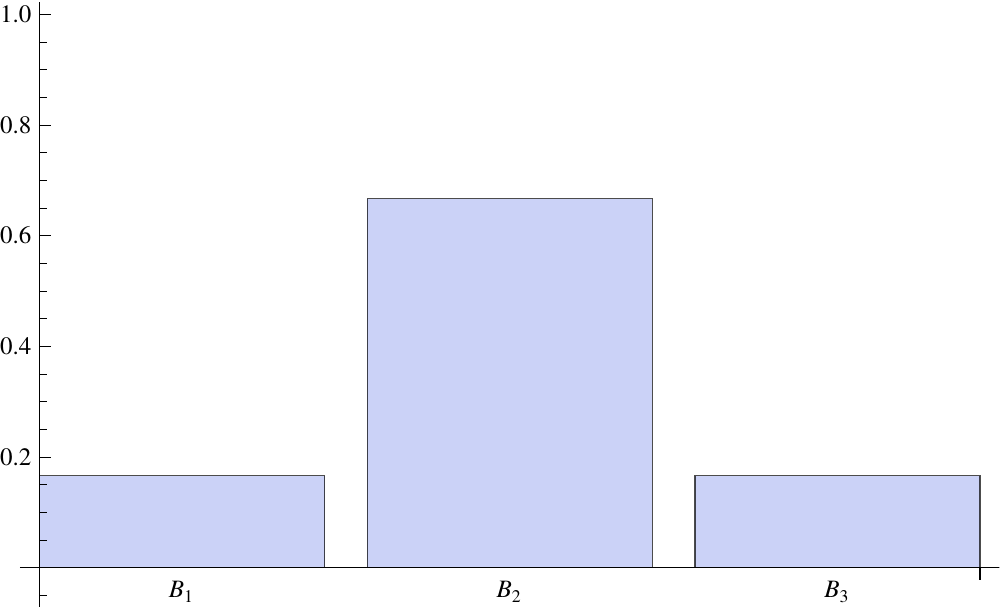} \\
\includegraphics[width=5cm]{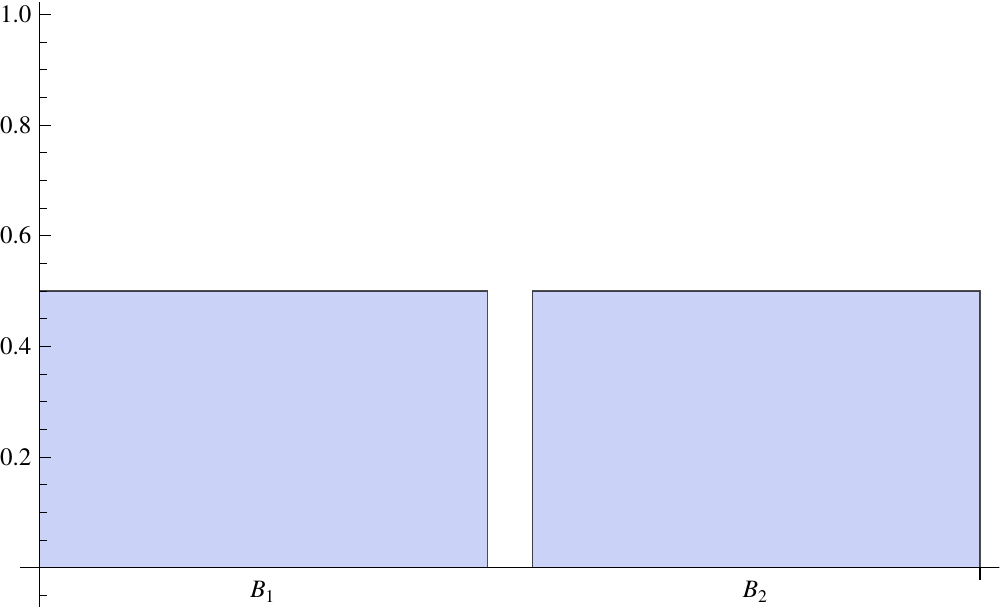} 
& \rule{0.5cm}{0cm} & \includegraphics[width=5cm]{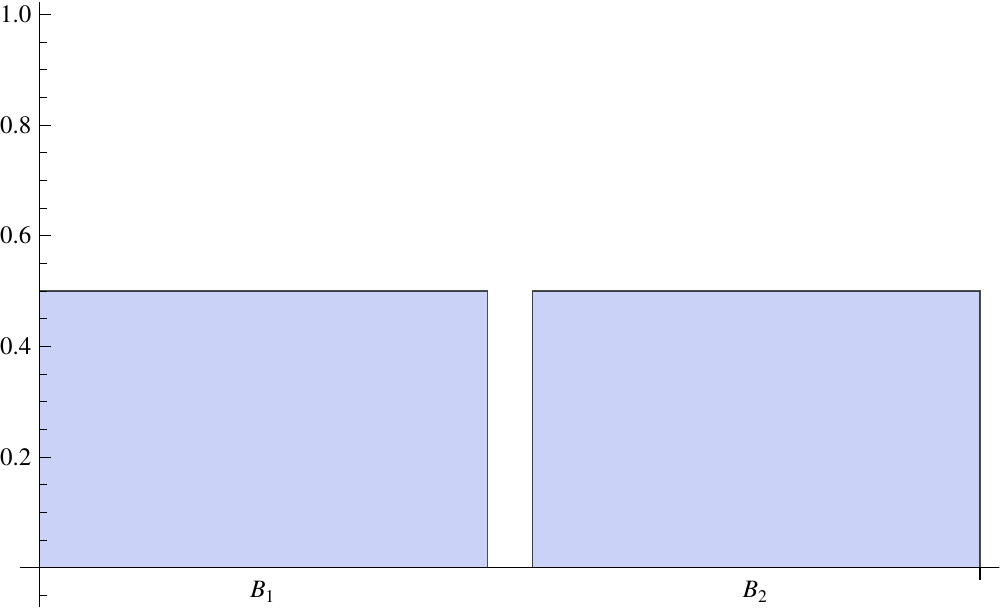}
\end{tabular}
\end{center}
\caption{\label{fig:compare_buckets}Bucketized distributions of the confidential
attribute for the whole data (left) and for an equivalence class $E_i$ (right).
Three buckets are considered in the top distributions, 
and two in the bottom ones.}
\end{figure*}

When bucketizing the distributions in Figure~\ref{fig:2_close_distribution},
the subrange with density 2 should exactly
correspond to a bucket or a union of buckets, in order
to maximize the utility of the resulting data. This is illustrated
in Figure~\ref{fig:compare_buckets}, whose top row shows
bucketized versions of the distributions 
of Figure~\ref{fig:2_close_distribution} using {\em three} buckets:
top left graph, bucketized version of the global distribution;
top right graph, bucketized version of the restriction to $P_i$.
Note that, for each of the buckets, the global probability
and the probability restricted to $E_i$ differ by a multiplicative
factor of two; that is, we attain $2$-closeness with equality for each
of the buckets. The bottom row of 
Figure~\ref{fig:compare_buckets} 
shows the bucketized versions of the
distributions in Figure~\ref{fig:2_close_distribution}
using {\em two} buckets. It can be seen that, with the two proposed buckets,
both bucketized distributions are identical; 
that is, we get $1$-closeness, which is
stronger than the intended $2$-closeness, but comes at the cost
of data utility loss.
Therefore, the number and hence the 
probability mass of the optimal buckets is dependent
on the level of $t$-closeness that we want. 

According to the previous example, if the privacy requirement is $t$-closeness,
for a certain $t$, it seems reasonable to use up the allowed distance $t$ between 
the global distribution of the confidential attribute and the restriction of that 
distribution within each equivalence class. Using up the allowed distance between the 
confidential attribute distributions enables forming equivalence classes that are more 
homogeneous, and hence decreases information loss.
Let us now put together what we said about the homogeneity and size of the buckets, and
about the equivalence classes that emphasize a specific bucket (the probability 
distribution of the restriction to each equivalence class must differ from the global distribution
by a factor of $t$ for a specific bucket, and by a factor of $1/t$ for the rest of buckets).
The conclusion is that we should set off for the maximum number of buckets (thus, for buckets that
are as small as possible) but with the constraint that equivalence classes should be able to
emphasize a specific bucket. This is formulated as
\[ t \times b_i/N + \frac{1}{t} \times (1-b_i/N) = 1 \]
for all $i$. The result is $b_i = N/(t+1)$ for all $i$. That is, the optimal bucketization
consists of $t+1$ buckets of size $N/(t+1)$. Notice that the optimal bucketization is only
attainable if $t+1$ divides $N$; otherwise we should select the closest feasible bucketization.

%

\subsection{$t$-Closeness construction}

Consider the original data set $X=\{(qi_{i},c_{i})|i=1,2, \cdots,N\}$,
where $qi_{i}$ refers to the quasi-identifier attributes, and $c_{i}$
to the confidential attribute. We want to generate a $k$-anonymous
$t$-close data set $X'$. 

According to Section~\ref{sub:Bucketizing}, we need to reduce the granularity
of the confidential attribute. In particular, it was proposed to group
the values of the confidential attribute in $t+1$ buckets of $N/(t+1)$.
records. In general a clustering algorithm is used to generate a set of buckets
that maximize intra-bucket homogeneity. Here, for the sake of simplicity, we assume 
that the records can be ordered in terms of the confidential attribute $c_{i}$ 
(this is possible if $c_i$ is numerical or ordinal). In this 
case the buckets are: 
\[
\begin{array}{c}
B_{1}=\{c_{1},c_{2},\cdots,c_{\left[\frac{N}{t+1}+0.5\right]}\},\\
B_{2}=\{c_{\left[\frac{N}{t+1}+0.5\right]+1},
c_{\left[\frac{N}{t+1}+0.5\right]+2},
\cdots,c_{\left[2\times\frac{N}{t+1}+0.5\right]}\},\\
\vdots\\
B_{t+1}=\{c_{\left[t\times\frac{N}{t+1}+0.5\right]+1},
c_{\left[t\times\frac{N}{t+1}+0.5\right]+2},
\cdots,c_{N}\}.
\end{array}
\]
The $k$-anonymous $t$-close data set is generated as follows: 
\begin{enumerate}
\item Replace the values of the confidential attribute
in the original data set $D$ by the corresponding buckets,
and call $\bar{D}$ the resulting data set;
\item Partition $\bar{D}$ in equivalence classes of $k$ (or more) records.
\end{enumerate}

In the second step above, not all values of $k$ are
equally suitable. For instance, it must be $k \geq t+1$,
because we showed in Section~\ref{sub:Bucketizing}
that $b \leq k$ and $b=t+1$.
In fact, we can write:
\[
k=\frac{N}{(t+1)l}
\]
where $l\ge1$ is a natural number that counts the number of equivalence classes 
that emphasize each of the buckets.
In fact, if we take into account
the previous inequality $k\geq t+1$, we conclude that $l$ belongs
to the set $\{1, 2,\cdots,\left\lfloor \frac{N}{(t+1)^{2}}\right\rfloor \}$.
Similarly to the discretization of the confidential attribute, the
value of $k$ produced by the previous formula may not be an integer. In
that case we need to adjust the size $e_{i}$ of each equivalence class $E_{i}$ to
\[ e_{i}=\left[i\frac{N}{(t+1)l}\right]-\left[(i-1)\frac{N}{(t+1)l}\right]. \]

\begin{table}
\caption{\label{tab:probabilities}Theoretical probability mass of the distribution
of the confidential attribute in each of the buckets corresponding
to the discretization of the confidential attribute. }

\begin{centering}
\begin{tabular}{ccccc}
 & $B_{1}$ & $B_{2}$ & $\cdots$ & $B_{t+1}$\tabularnewline
\hline 
Original data & $\nicefrac{1}{t+1}$ & $\nicefrac{1}{t+1}$ & $\cdots$ & $\nicefrac{1}{t+1}$\tabularnewline
\hline 
$E_{1}$ & $\nicefrac{t}{t+1}$ & $\nicefrac{1}{t(t+1)}$ & $\cdots$ & $\nicefrac{1}{t(t+1)}$\tabularnewline
$E_{2}$ & $\nicefrac{1}{t(t+1)}$ & $\nicefrac{t}{t+1}$ & $\cdots$ & $\nicefrac{1}{t(t+1)}$\tabularnewline
$\vdots$ & $\vdots$ & $\vdots$ &  & $\vdots$\tabularnewline
$E_{t+1}$ & $\nicefrac{1}{t(t+1)}$ & $\nicefrac{1}{t(t+1)}$ & $\cdots$ & $\nicefrac{t}{t+1}$\tabularnewline
\hline 
\end{tabular}
\par\end{centering}

\end{table}

Table~\ref{tab:probabilities} gives the theoretical probability
mass of 
each bucket of the confidential attribute for each of the equivalence classes.
We assume that $l=1$ and that equivalence class $E_{1}$
emphasizes bucket $B_{1}$, $E_{2}$ emphasizes bucket $B_{2}$, and
so on.
The exact theoretical probability masses
may not be achievable due to the discrete nature of the data. First
of all, it may not be possible to obtain a discretization of the confidential
attribute in buckets with probability mass $1/(t+1)$.
Also, when generating the $k$-anonymous partition $E_{1}, E_{2},\cdots,E_{t+1}$,
it may not be possible for each of the groups to contain
exactly $k$ records. Let $p_j=b_j/N$ be the probability that a record in the original data set
belongs to bucket $B_j$.
For $t$-closeness to be achieved, the following must 
hold for every equivalence class $P_i$: (i)
at most $\left\lfloor e_{i}p_{i}t\right\rfloor $ records must have
$B_{i}$ as the value for the confidential attribute; and (ii) at
least $\left\lceil e_{i}p_{j}/t\right\rceil $ records must have $B_{j}$
as the value for the
confidential attribute. For these conditions to hold, we can start
selecting $\left\lceil e_{i}p_{j}/t\right\rceil $ records with confidential
attribute $B_{j}$, for each $j\ne i$, and complete the partition
set with $e_{i}-t\left\lceil e_{i}p_{j}/t\right\rceil $ records with
confidential attribute $B_{i}$.

\subsection{$t$-Closeness algorithm}
Let us now restate the whole bucketization and equivalence class generation process 
in an algorithmic way:
\begin{enumerate}
\item Let the number of records in the original data set be $N$.
\item Let $t$ be the desired level of $t$-closeness.
\item Cluster the $N$ values of the confidential attribute in 
the original data set into $b=t+1$ buckets in such a way that:
\begin{enumerate}
\item the probability mass of each bucket 
is as close as possible to $1/b$, that is,
each bucket contains $[N/b]$ values, except for some buckets that contain
$[N/b]+1$ values (when $N$ is not divisible by $b$);
\item values within a bucket are as similar as possible.
{\em E.g.} for a numerical or ordinal confidential attribute, 
each bucket would contain consecutive values.
In general, a clustering algorithm can be used.
\end{enumerate}
In this way, we can 
view the bucketized distribution of the confidential
attribute in the original data set as being
uniform.
\item Partition the records in the bucketized data set 
into a number of equivalence classes,
in such a way that every equivalence class satisfies that:
\begin{enumerate}
\item it contains $k$ (or more) records,
in view of achieving $k$-anonymity;
\item  no bucket contains a proportion
of the confidential attribute values of the equivalence class
higher than $t/b$ or lower than $1/(tb)$ (that is,
so that the bucketized distribution of the confidential attribute
in the group is at distance less than $t$ from
the bucketized distribution of the confidential 
attribute in the overall data set, according to 
Definition~\ref{def:distance}).
\end{enumerate}
\end{enumerate}

\section{Related Work}
\label{related}

Our proposal aims to find links between syntactic privacy models (in particular, $t$-closeness) and differential
privacy. Syntactic privacy models require the anonymized data set to have a specific form that helps reducing the
disclosure risk. $k$-Anonymity~\cite{SamaratiSweeney98,Samarati01}, $l$-diversity~\cite{ldiver}, and 
$t$-closeness~\cite{tclose,tclose2} are among the most popular syntactic privacy models. These privacy models are
based on specific intruder scenarios, and they aim at avoiding data set configurations that are disclosive under
these intruder scenarios. Syntactic models of privacy are known to have several issues such as their limited utility
for high-dimensional data sets~\cite{Aggarwal2005} and the vulnerabilities they present against several 
attacks~\cite{Chi07,ldiver,tclose}. Perhaps the most prominent issue with syntactic privacy models has to do
with the intruder scenario: if the level of knowledge of the intruder is greater than assumed, the protection achieved
may be ineffective.

Differential privacy~\cite{Dwork06,Dwork06b} was introduced 
to provide a strong privacy model that
addresses the vulnerabilities of previous privacy models. To this end, differential privacy takes a relative
approach to disclosure limitation: the risk of disclosure must be only slightly affected by the inclusion or 
removal of any specific record in the data set. 
In this way, differential privacy avoids the need to make
assumptions about intruder scenarios
(an intruder that knows everything but one record is implicitly assumed).

The dissimilar approaches to disclosure risk limitation 
taken by differential privacy on the one hand 
and syntactic models on the other hand
have motivated mutual criticism between both
families of methods. 
For example, \cite{Dwork07,Dwork08,Dwork11} justify the use of differential 
privacy by criticizing several statistical
disclosure control techniques such as query restriction, input perturbation 
and even output perturbation (the basis of differential privacy) 
when applied naively. On the contrary~\cite{Muralidhar10,Bambauer14} 
criticize differential privacy because of the limited utility it provides for numerical data. Other criticisms to
differential privacy are related to the unboundedness of the responses and to the selection of the $\varepsilon$ 
parameter. See~\cite{Clifton13} for a more detailed comparison between syntactic privacy models and differential
privacy.

Despite the above controversies, 
some attempts to find connections between differential privacy and syntactic 
privacy models have been made. In~\cite{Li14} it is shown that when $k$-anonymization is done ``safely''  it leads
to $(\varepsilon,\delta)$-differential privacy (a generalization 
of differential privacy). The randomness required for
differential privacy to hold is introduced by a random sampling step. In contrast to~\cite{Li14}, which only goes
from $k$-anonymity to $(\varepsilon,\delta)$-differential privacy, we show both implications between 
$\varepsilon$-differential privacy and $t$-closeness. The approach used to introduce the uncertainty required
by $\varepsilon$-differential privacy is also different; while~\cite{Li14} make use of an additional sampling
step, we take advantage of the assumptions made
by $t$-closeness about the prior and posterior views of the data.
Connections between syntactic models of privacy and 
differential privacy are not limited to the satisfiability
of one family given the other. In~\cite{Gehrke12} an interesting mix between differential privacy and $k$-anonymity is
proposed. Essentially, $\varepsilon$-differential privacy 
is relaxed to require individuals to be indistinguishable
only among groups of $k$ individuals. In~\cite{soria14} 
$k$-anonymity is used as an intermediate step in the
generation of an $\varepsilon$-differentially private data set. The use of $k$-anonymity reduces the sensitivity
of the data, thereby decreasing the amount of noise 
required to satisfy differential privacy.

\section{Conclusions and future research}
\label{sec7}

This paper has highlighted and exploited several
connections between $k$-anonymity, $t$-closeness
and $\varepsilon$-dif\-ferent\-ial privacy. These models
are more related than believed so far in the case 
of data set anonymization.

On the one hand we 
have introduced the concept of stochastic $t$-closeness,
which, instead of being based on
the empirical distribution like
classic $t$-closeness,  is based on the distribution
induced by a stochastic function that modifies the confidential attributes. 
We have shown that
$k$-anonymity for the quasi-identifiers combined with 
$\varepsilon$-differential privacy for the confidential 
attributes yields stochastic $t$-closeness,
with $t$ a function of 
$\varepsilon$,
the size of the data set
 and the size of the equivalence classes.
This result shows that differential privacy is stronger than $t$-closeness
as a privacy notion. From a practical point of view, it provides
a way of generating an anonymized data set that satisfies both (stochastic)
$t$-closeness and differential privacy.

On the other hand,
we have demonstrated that the $k$-anonymity family of models 
is powerful enough to achieve $\varepsilon$-differential 
privacy in the context of data set anonymization, 
provided that a few reasonable assumptions on the intruder's
side knowledge hold. 
Specifically, using a suitable construction,
we have shown that $\exp(\varepsilon/2)$-closeness
implies $\varepsilon$-differential privacy.
The construction of a $t$-close data set based on
the distance function in Definition~\ref{def:distance} has also
been detailed. Apart from partitioning into equivalence classes,
a prior bucketization of the values of the
confidential attribute is required. The optimal size of 
the buckets and the optimal size of
equivalence classes have been determined.
The new stochastic $t$-closeness model opens several future research lines.
Being a generalization of $t$-closeness, we can expect stochastic $t$-closeness 
to allow better data utility. Comparing the utility obtainable 
with both types of $t$-closeness is an interesting future research line 
that requires devising a construction to reach stochastic $t$-closeness
(other than the one based on differential privacy). 
Exploring whether and how stochastic
$t$-closeness (rather than standard
$t$-closeness) could yield $\varepsilon$-differential privacy 
is another possible follow-up of this article.
Finally, it would also be interesting to compare in
terms of privacy and utility the impact
of the distance between distributions proposed in the article and the earth mover's distance.

%

\section*{Acknowledgments and disclaimer}

The following funding sources are gratefully acknowledged:
Government of Catalonia (ICREA Acad\`emia Prize to the
first author and grant 2014 SGR 537),
Spanish Government (project TIN2011-27076-C03-01 ``CO-PRIVACY''),
European Commission (projects
FP7 ``DwB'', FP7 ``Inter-Trust'' and H2020 ``CLARUS''),
Templeton World Charity
Foundation (grant TWCF0095/AB60 ``CO-UTILITY'')
and Google (Faculty Research Award to the first author).
The authors are with the UNESCO Chair in Data Privacy.
The views in this paper are the authors' own and
do not necessarily reflect
the views of UNESCO, the Templeton World Charity
Foundation or Google.

\end{document}